# Multidimensional Analysis of Psychological Factors affecting Students' Academic Performance

Leena Khanna
ASET, Amity University,Uttar Pradesh, 201313, India

Shailendra Narayan Singh
ASET, Amity University, Uttar Pradesh, 201313, India

Mansaf Alam
Department of Computer Science, Jamia Millia Islamia, New Delhi, 110025, India

**Abstract**
Academic performance of any individual is dependent upon numerous aspects regarding the day to day life of the individual under consideration. Academic performance is measured in terms of the grade point average or GPA as it is called. Grade point average is dependent not only on the faculty but also on various psychological parameters including the study habits, social anxiety and allied. In this study, a detail analysis of numerous psychological factors impacting the grade point was carried and based upon various psychological factors the performance for the student in forth coming examination was forecasted.

**Keywords**
Forecasting, Grade Point Average, Achievement Motivation Scale, Study Habits, extraversion, conscientiousness, social anxiety, emotional competence, self-efficacy for children.

## 1. Introduction

In the modern era, the outcome of academic initiatives which are initiated by parents or the academic institutes depends upon various factors. These factors vary a great extent ranging depending upon the psychological condition of an individual. A change in any of the psychological indicator can increase or decrease the performance of a student in any examination. This in turn will have impact on the career of the student as well as the annual result of the institution. Thus, it is a prime concern for any education institute to map the psychological wellbeing of a student with the performance (w.r.t. Grade Point Average (GPA)).

The individual differences which have an impact on the grade point equivalent have been extensively researched. There had been numerous studies which tried to find out the various psychological parameters impacting the final outcome on grade point average. One highly important study was done by Rothstein, et. al. in 1994 [1] where they found that there are rational and realistic ground that the academic performance of an individual can be forecasted using the various personality traits of the individual.

In the present scenario, where the dynamics of the global job markets are changing every day, the requirement of quality work force coming from higher educational institutions is extremely high. With the advent of technology and especially, the advancement in the field of artificial intelligence has resulted in the various tools being available for analysing the psychological status of a student and thus, depicting the impact of this over his grade point average.

The research conducted by various academic institutions in the field of data mining has contributed a great deal of knowledge in the domain of forecasting technologies. The utilization of the any data mining technique is based upon the fact that the data holds essential features which can be utilized for predicting the future with a certain amount of certainty and thus, assist in decision making.

In this paper, we put present a method which can analyse the impact various psychological factors impacting the performance of a student in higher education institutions. The design of system incorporates the use of various psychometric scales in harmony with different techniques from data mining to generate the grade point average outcome of the student. The inference system depicts the approach taken by an expert clinical psychologist. The final forecast generated by the system is designed to help the academicians to help the institutes in focusing upon the individual students who because of various psychological parameters have a decreased performance in the examination and thus, increasing the overall institution results as well as help students fight against various psychological issues & aid in the counselling of students.

The paper is structured in the 7 sections. In section 2, we are presenting the literature review of the domain in consideration. In section 3, we're explaining various psychological scales which are used in this study. In section 4 and 5, the various data mining techniques and research methodology is discussed, respectively. Finally in section 6, we put forward the result of the study. In section 7, we conclude the study.

## 2. Literature Review

There is widespread use of data mining in the higher education system. Work has been carried by numerous researchers in this field. They have gone through various aspects of data mining in the domain of education. In this section we present a holistic survey of various psychological as well as data mining techniques as existing in the research domain and the methodology followed to complete the literature survey.

### 2.1. Literature Review Procedure

The framework for conducting the literature review is separated in 3 parts: scope of survey, identification of research literature and Reporting the literature of interest and is depicted in figure 1.

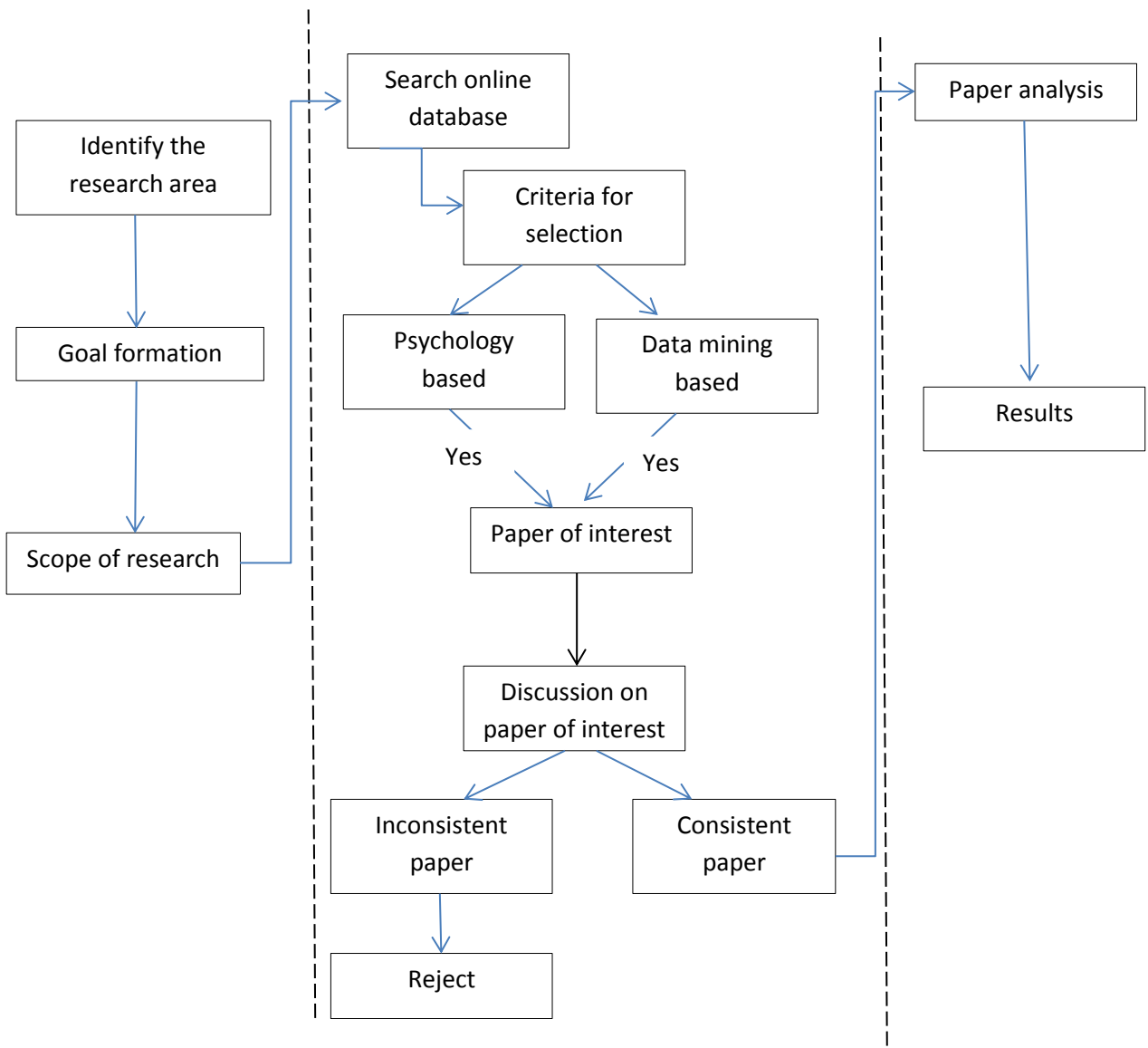

Figure 1: Procedure for literature review

In part 1, we have defined the problem under consideration. This phase has been divided into 3 phases. In this phase, the study starts by describing the area of the research which is then followed by the formulation a research goal describing the process of the survey for the literature review. This process ends up by producing a unique statement of problem which becomes the basis upon which the future work will be conducted. The domain of research happens in this study to be the psychological parameters, the data mining techniques and their intersection over higher educational domain. The aim of this research work is to design a method in which can be utilized by higher educational institutions to help students perform better in academics.

In part 2, we have framed the conditions for choosing of research literature pertaining to the domain of study. Seven online databases were examined to figure out the details for work being done in domain of study. These databases are:

- Elsevier
- IEEE
- Taylor &Francis
- ACM
- Oxford
- Science Direct
- Springer

The research databases were searched using the keywords "Forecasting", "Grade Point Average", "Achievement Motivation Scale", "Study Habits", "extraversion", "conscientiousness", "social anxiety", "emotional competence", "self-efficacy for children". To search for different permutations of the keywords Boolean operators were used.

## 2.2 Classification of under research literature

This section provides a holistic view over the research literature being considered in this study. A total 44 research papers were consulted during this study. This review of this research literature is classified in this section owing to the findings, year as well as the major data mining technique which has been used in paper. These are as follows:

### 2.2.1. Classification based upon the major finding

This section presents the major finding/designed developed mapped to the research literature in table 1.

| S. No. | Reference number | Major findings |
|---|---|---|
| 1 | [1] | Method for literature survey |
| 2 | [2] | Representation of institutional growth perspective based upon the performance of students. |
| 3 | [3] | Analysed the students data and generated clusters of strong, medium and weak performance |
| 4 | [4] | Detected the set of weak students |
| 5 | [5] | Designed a predicting model for performance in academics |
| 6 | [6] | Detected the outlier of low academic achievers |
| 7 | [7] | Designed a comprehensive style of student evaluation |
| 8 | [8] | Forecasted the chances of success of curricula |
| 9 | [9] | Student performance analysis |
| 10 | [10] | Detected the outlier of weak students |
| 11 | [11] | Mapped attributes of enrolment with the academic success |
| 12 | [12] | Examined the academic accomplishment of students to their |

| | | financial growth |
|---|---|---|
| 13 | [13] | Drop out forecasting |
| 14 | [14] | Analysed e-learning success rate |
| 15 | [15] | Designed a system to classify the grades |
| 16 | [16] | Evaluated the effectiveness of study in students |
| 17 | [17] | Designed prediction model to predict the performance on the basis of behaviour. |
| 18 | [18] | Forecasted achievements of students |
| 19 | [19] | Improvised apriori to mine data tables |
| 20 | [20] | Detected students weak in studies |
| 21 | [21] | Found out the factors impacting the success as well as failure among students |
| 22 | [22] | Forecasted the final grade depending upon the historical performance |
| 23 | [23] | Socio-economic variables can influence the academic outcome from students |
| 24 | [24] | Designed system to predict the final grades |
| 25 | [25] | Academic performance classification |
| 26 | [26] | Outlier detection of weak students |
| 27 | [27] | Gave the parameters for the analysis of performance in academic institutes |
| 28 | [28] | forecasted the final grade |
| 29 | [29] | Predicted the performance of an examination |
| 30 | [30] | Comparison of algorithms in terms of forecasting accuracy |
| 31 | [31] | Comparison of algorithms in terms of forecasting accuracy |
| 32 | [32] | Comparison of algorithms in terms of forecasting accuracy and concluded that classifier is best for classification |
| 33 | [33] | Comparison of algorithms in terms of forecasting accuracy |
| 34 | [34] | Graduates lack critical thinking, team work and analytical approach |
| 35 | [35] | Comparison of algorithms in terms of forecasting accuracy |
| 36 | [36] | Designed a satisfaction criterion for a course under consideration |
| 37 | [37] | Evaluation of faculty staff |
| 38 | [38] | System for assistance in enrolment |
| 39 | [39] | Found out that multi-lingual class is more suited than unilingual class |
| 40 | [40] | Forecasted the learning disabilities |
| 41 | [41] | Patterns in matching associations |
| 42 | [42] | Predicted the sequence for course registration |
| 43 | [43] | Analysis of continuation of education |
| 44 | [44] | Prediction of enrolment head count |
| 45 | [46] | Educational Data Mining and its Role and Impact on Society |

| 47 | [47] | Use of Genetic Algorithm to optimize features affecting the Grade performance Average of Students |
| 48 | [48] | Multiple Filter and Wrapper Approach are used to identify the parameters affecting the Grade performance Average of Students |

Table 1 : Classification based upon key findings

### 2.2.3. Classification based upon the Data mining techniques used

To figure out the key algorithms utilized for forecasting in the domain of higher educational research, we have done an analysis of higher education and the data mining techniques. Eleven data mining techniques have been utilized in the analysis of higher educational domain. It is clear form this analysis presented in table 2 that decision tress was the most widely used technique in this domain followed by the Naive Bayesian algorithm.

| S.No. | Data Mining Technique | Reference number |
|---|---|---|
| 1 | Naive Bayes | [2],[5],[6],[23],[24],[29],[30],[35],[42],[45],[48] |
| 2 | K-means Clustering | [3],[9],[10],[12],[27],[31],[32],[34] |
| 3 | Association rule mining | [4],[7],[12],[18],[21],[39],[41] |
| 4 | Decision Trees | [5],[8],[11],[12],[14],[15],[16],[17],[24],[26],[27],[29],[30],[31],[32],[33],[34],[35],[36],[38],[40],[42],[43] |
| 5 | Artificial Neural Networks | [9],[13],[29],[32],[33],[36],[37] |
| 6 | Rough Set Theory | [13],[20],[24] |
| 7 | Apriori Algorithm | [19] |
| 8 | Genetic algorithms | [22],[28],[47] |
| 9 | Fuzzy logic | [25],[33] |
| 10 | Support vector machine | [36] |
| 11 | Random Forest | [37], [44] |

Table 2 : Classification based upon the technique used

### 2.2.4. Classification based upon the year of publication

Figure 2 depicts the spread of articles referred in this study as per their year of publication.

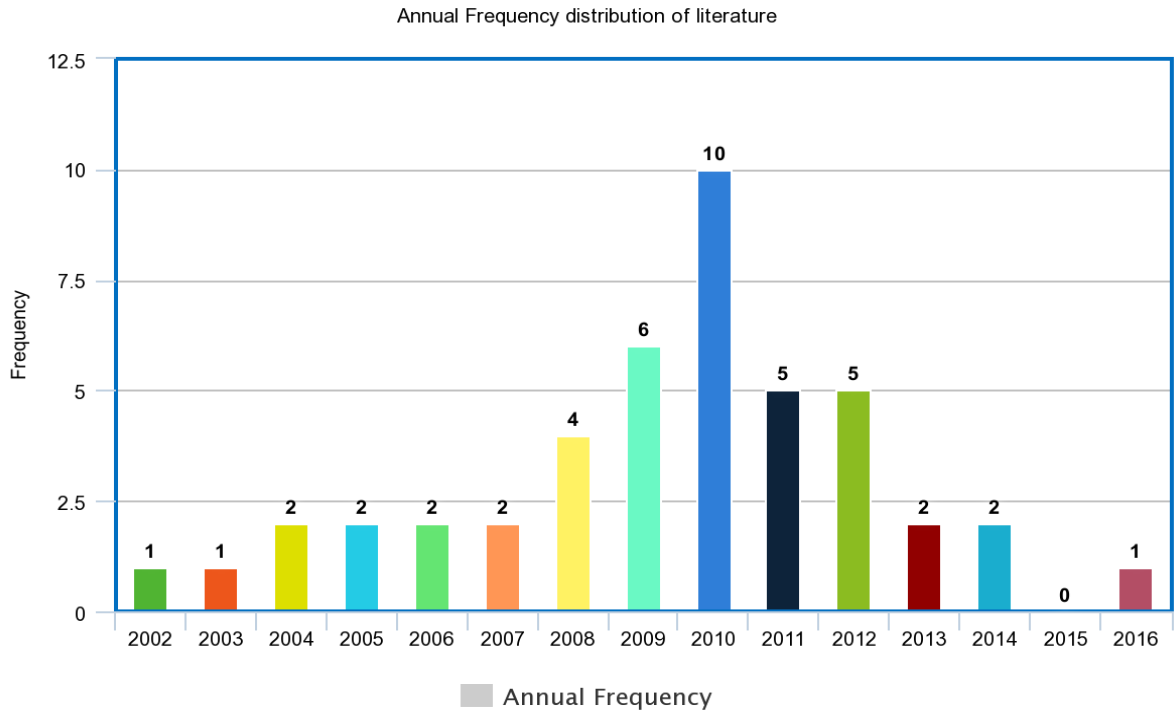

Figure 2: Classification based upon the year of publication

## 2.3. Forecasting Overview of Data Mining for psychological analysis

In the history of human civilization forecasting has been done since ages. In this section, we classify the existing work in the domain of consideration into broad themes of the data mining and its applicable techniques. The classification as depicted in the figure 2 shows the interaction of psychological approach with the data mining technique. Psychological analysis is a highly interdependent part thus sits in the center of analysis.

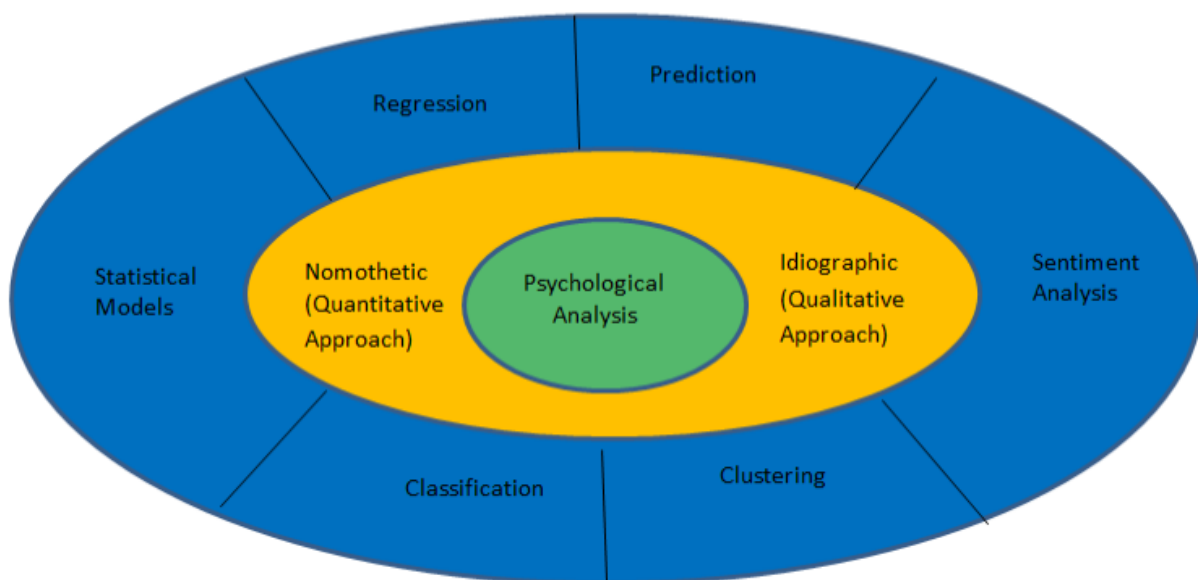

Figure 3: Forecasting framework of data mining for psychological analysis

The proposed overview of data mining for psychological analysis is based upon the theoretical literature available in the domain under consideration.

### 2.3.2. Psychological Analysis

In psychology the analysis is done in two different ways which are as follows:

### 2.3.2.1 Nomothetic (Quantitative approach)

The word "nomothetic" is derived from the Greek word "nomos" which translates to law. This is utilised by the psychologists while trying to study what the subject under consideration shares with other people around him. This approach is used for the analysis and interpretation of data gathered by an investigator incorporating the use of inferential as well as descriptive statistics. This is utilized while incorporating the use of a psychological test.

### 2.3.2.2. Idiographic (Qualitative Approach)

The word "idiographic" is derived from the Greek word "idios" which means "private". Psychologists who are interested in this aspect of study seek the experience of each individual which makes them unique. This approach is dependent upon the qualitative methods of collecting the data such as diaries, interview and other allied written materials which have either been obtained from the subject under consideration or is expected to have been belonging to the same.

## 3. Psychological scales

In this study we are using a total of 7 psychological scales were used which depict various attributes of personality. These were selected on the basis of various personality traits which were under consideration and had shown influence over academic performance in the research literature. These personality traits are as follow:

### 3.1. Achievement Motivation

It is a type of psychological test designed to evaluate a wide set of work-related achievement motivation. In this study, it is being used to find out the motivation to achieve a higher competence in academic performance. In this work, the achievement motivation scale by Prathiba Deo and Asha Mohan was used.

### 3.2. Study Habits

These are the sum total of all the practices that are utilized by any individual to learn any aspect of life or subject. Study skills are mostly defined as the ability of the student to manage all the resources including time to full-fill an academic task with high amount of certainty. Thus, study habits happen to be the different forms of routines which are followed by a student during any period in a conductive environment. In this study, study habit inventory by M.N. Palsane and Anuradha Sharma are utilised.

### 3.3. Extraversion

It is defined as a state in which one obtains gratification not from the inside but from outside oneself. People with this trait enjoy human communications and tend to be talkative, enthusiastic as well as assertive. These people are always energized and take pleasure in activities that incorporate large number of people. These people also tend to work best in a group. In this study, Extraversion as defined in L.R. IPIP scale measures in utilized.

### 3.4. Conscientiousness

It is the personality trait which depicts how careful/vigilant a person is. The word "conscientiousness" itself means an aspiration to perform a task well and take into serious consideration the obligations by others. People with this personality trait tend to perform more in terms of efficiency and are highly organised. They depict an inclination to display self-discipline, dutiful acts as well as target for achievement in their respective field. They display a planned behaviour and are generally reliable. In this study, the conscientiousness as defined in Goldberg IPIP scale measures is used.

### 3.5. Social Anxiety

Social anxiety is the nervousness faced by some people in social situations. Individuals which are in social anxiety depict behaviour of avoiding interaction in form of gazes, lesser number of facial expressions as well as difficulty in starting and maintaining conversations. Most of the people have faced social anxiety at some point of time in their life. In this study, Social Anxiety scale named as SIAS is used.

### 3.6. Emotional Competence

Emotional Competence is one's capability to show his/her inner feelings to others in a comprehensive form. This shows the easiness of a person when around other people to express. It is one of the essential social skills to identify, understand and react to emotions of yourself as well as others around you. In this study, the emotional competence scale by Harish Sharma and Rajeev Lochan Bhardwaj is utilized.

### 3.7. Self- Efficiency

Self-efficiency is the belief of an individual in his/her inborn abilities to achieve the desired goals. Psychologists have been studying self-efficiency from numerous perspectives. But in this study, we stick to the utilization of self-efficiency to the domain of achievement in higher education. To deduce this self-efficiency questionnaire SEQ-C was utilized.

### 4. Data Mining Techniques

Data mining is the analysis of existing data to find out new information out it. Usually, it is done over tremendous amount of data to figure out the required patterns of interest. The large scale implementation information technology has grown leap and bounds generating different systems for transactions and different ones for analytics. Data mining provides the bridge between the two. In this study, as eminent from the literature review, we'll be using the following methods to analyse the data. These methods are:

**4.1. Naive Bayes Classifier**

Naïve Bayes classifier belongs to the family of simple probabilistic classifier which utilises the Bayes theorem and implements a strong assumption of independence among the features being classified. These are highly saleable classifiers which need a linear number of variables to the predictors in the learning problem.

**4.2. Decision Tress**

As evident from the literature review, decision tree is the most commonly used technique in this domain. A decision tree happens to be a decision support tool which implements a graph structure and its leaf nodes depict the outcome and the intermediate nodes represents the state in which the machine is presently is. It follows a sequence of decisions to conclude a possible outcome. It happens to be one of the ways to depict an algorithm which has conditional control statements.

**4.3. K-means Clustering**

It is a method of quantizing the input vectors and is quite popular in the domain of cluster analysis. The aim of k-means is to divide the number of observations say n into a particular number of clusters say k and the observations belong to one of the k clusters depending upon their mean. The partitions generated in this process are known as voronoi cells. As per table 2, it has been widely used in the research literature.

**4.4. Artificial Neural Networks**

These are a set of computational methods which are inspired from the biological neural networks which make up the brain of animal brains. These are designed to learn the anticipated behaviour from the examples which are provided to them. These are deployed in used in the research literature for their power to forecast an unknown scenario based upon previous data. In this study, back propagation algorithm is used as the training method for feed forward network.

**5. Methodology**

This research is carried out in three phases. In the 1$^{st}$ phase, the various psychological questionnaires were arranged in a random order and were distributed in 14 institutions of higher education for data collection from graduate and post-graduate courses. In phase 2, the information was processed to give it a digital shape and noise was removed. In the 3 phase, these were analysed using the techniques as mentioned in the section 4 of this paper and present the achieve results. In details, these phases are as follows:

**5.1. Data Collection**

This was one of the initial phases of the research. Once the psychometric scales and the questionnaire were decided, the final set of questions was distributed across 14 institutions of higher education and was filled by over 1000 students from the graduation as well as post-graduation courses. This was come in parts and was arranged and translated into the electronic format. Except for the psychological scales, the grade point average of students was collected along with their type of course.

**5.2. Data Pre-processing**

As soon as the data collection was completed, it was noticed that few students has not filled the questionnaire fully. These entries were deleted from the data set. This bought the effective number down to 837.

The data collected over the psychometric scales was arranged and then min-max normalization was utilised over psychometric data. This helped us to range bound the data and thus, helped range bound the data improving the final outcome of the study.

The grade point equivalent followed in different educational institutions was different. Few instructions followed 10-point scale while some followed 4-point scale. To bring both on level plane field and comparable, the 10-point scale was reduced to 4-point scale. Thus, all the entries in the database were translated in a 4-point scale which ranged from 0-4. Since, these are defined as grade which ranged from A to F, we arranged these grades into clusters of EXCELLENT, GOOD, AVERAGE and BAD. The range of vales these belongs is as in table 3.

| S. No. | Range | Name |
|---|---|---|
| 1 | 3.67-4 | Excellent |
| 2 | 2.67-3.67 | Good |
| 3 | 1.67-2.67 | Average |
| 4 | 0-1.67 | Bad |

Table 3: nomenclature of clusters

**5.3 Achieved Results and Discussion**

The main objective of this research was to predict the class in which a student will belong in his coming examination based upon various psychological inputs which are retained in the model. Several different models as explained in the section 4 of this paper are applied to the dataset to build a classification model and a final comparison of all these models is provided. To build these models MATLAB was used. The data was split into training data, testing data and validation data. The ratio of the split was 7:1.5:1.5 i.e. 70% of data was used for training, 15% each for testing and validation of the trained model. To present the results, in this section we'll be using true positive rate and the precision. True positive rate is the percentage of examples which are classified to their respective class whereas precision is defined as all the examples which are mapped to a class and they belonged to that class as well.

### 5.3.1. Decision Tree Classifier

These are very powerful as well as popular tools in the domain of classification. These happen to be a tree like structure which starts with a starting node belonging to the root attributes and finish their journey at the leaf node depicting a class. In our case, the input vector was mapped to four classes. Thus, there happen to be 4 leaf nodes representing 4 classes. The advantage of the decision tresses is this that generates and represents rules in their intermediate nodes which are easy to understand and interpret by a human user.

The data was fed into MATLAB work space and decision tress was implemented over the same. The results are presented in the table 4.

| S. No. | Predicted class | True positive rate | Precision |
|---|---|---|---|
| 1 | Excellent | 0.89 | 0.91 |
| 2 | Good | 0.83 | 0.52 |
| 3 | Average | 0.63 | 0.78 |
| 4 | Bad | 0.81 | 0.67 |
| 5 | Average | 0.79 | 0.72 |

**Table 4: Decision Tree Classifier Results**

### 5.3.2. Naïve Bayes Classifier

This is a class of statistical classifier based upon the Bayes theorem which predicts the class membership by utilizing the probabilities of different sets involved. This algorithm assumes that the effect an attribute produces over a class is independent of the attributes. In other words, this means that in the input vector, the impact of each unit is different and is independent of each other. Thus, these are graphical models which can describe the joint conditional probability of the different elements in the input vector. These are known for high accuracy in certain domains like text analysis. The results of the analysis are provided in the table 5.

| S. No. | Predicted class | True positive rate | Precision |
|---|---|---|---|
| 1 | Excellent | 0.523 | 0.325 |
| 2 | Good | 0.412 | 0.345 |
| 3 | Average | 0.675 | 0.648 |
| 4 | Bad | 0.798 | 0.823 |
| 5 | Average | 0.602 | 0.53525 |

**Table 5: Naive Bayes classifier Results**

### 5.3.3. K-nearest Neighbour (K-NN) Classifier

This is a method of classification in which the training set is classified based upon the difference of vector from nearest matching cluster. It's a type of instance based learner which is also known as lazy learners. In our case, the numbers of clusters are 4. Thus, k in our case becomes 4 each cluster representing one of the desired states. The results are presented in table 6.

| S. No. | Predicted class | True positive rate | Precision |
|---|---|---|---|
| 1 | Excellent | 0.423 | 0.652 |

| 2 | Good | 0.78 | 0.714 |
| 3 | Average | 0.512 | 0.538 |
| 4 | Bad | 0.813 | 0.896 |
| 5 | Average | 0.632 | 0.7 |

Table 6 : K-NN classifier results

### 5.3.4. Artificial Neural Networks (ANN)

These are a biologically inspired set of algorithms which are widely used in the domain of machine learning. These are used clustering, classification and forecasting quite often. In our case, we are using these for classifying the input vector into one of the four desired classes. The architecture which provided local optima was 7-15-4. A feed-forward back propagation Neural Network was trained using this architecture and the results of the experiments are given in the table 7.

| S. No. | Predicted class | True positive rate | Precision |
|---|---|---|---|
| 1 | Excellent | 0.689 | 0.648 |
| 2 | Good | 0.712 | 0.735 |
| 3 | Average | 0.789 | 0.641 |
| 4 | Bad | 0.856 | 0.914 |
| 5 | Average | 0.7615 | 0.7345 |

Table 7: Artificial Neural Networks Results

### 5.3.5. Performance Comparison

The performances of all the techniques used in this study are compared in table 8. This this table shows that decision tree based classifier provided the results with the highest certainty.

| S. No. | | Decision Tree | K-NN | Naïve Bayes | ANN |
|---|---|---|---|---|---|
| 1 | True positive | 0.79 | 0.632 | 0.602 | 0.7615 |
| 2 | Precision | 0.72 | 0.7 | 0.53525 | 0.7345 |
| 3 | Overall | 0.755 | 0.666 | 0.568625 | 0.748 |

Table 8: Performance Comparison

## 6. Conclusion & Future work

In this paper, various Data Mining techniques are utilized. The study is consistent with the previous researches where it is identified that the accuracy of Decision Tree is better as compared to other techniques. In future, a general method to forecast the group a student belongs to in his forthcoming examination based upon numerous psychological factor will be proposed. This study helps to understand that psychological parameters can cause results of a student vary greatly. There are multiple factors which are taken into consideration to analyse their dependence on GPA, but which factor plays the most imperative role is yet to be studied. If these factors are taken into consideration well in time, the results of students as well as academic institutions will improve greatly and overall quality of Indian Educational System can be enhanced to a greater extent.